%% file: BLAST12_SPIE_arxiv.tex
\documentclass[]{spie}  
\usepackage[]{graphicx}
\usepackage{color}
\usepackage{amssymb}
\usepackage{wasysym}

\newcommand{\arcsec}{$^{\prime\prime}$}
\newcommand{\arcmin}{$^{\prime}$}

\title{The Balloon-borne Large Aperture Submillimeter Telescope for Polarimetry-BLASTPol: Performance and results from the 2012 Antarctic flight} 

\author{
N. Galitzki\supit{a},
P. A. R. Ade\supit{b},
F. E. Angil\`e\supit{a},
S. J. Benton\supit{c},
M. J. Devlin\supit{a}, 
B. Dober\supit{a},
L. M. Fissel\supit{c,d},
Y. Fukui\supit{e},
N. N. Gandilo\supit{c},
J. Klein\supit{a},
A. L. Korotkov\supit{f},
T. G. Matthews\supit{g}, 
L. Moncelsi\supit{h},
C. B. Netterfield\supit{c,i,j}, 
G. Novak\supit{g}, 
D. Nutter\supit{b},
E. Pascale\supit{b},
F. Poidevin\supit{k,l},
G. Savini\supit{m},
D. Scott\supit{d},
J. A. Shariff\supit{c},
J. D. Soler\supit{c,n}, 
C. E. Tucker\supit{b},
G. S. Tucker\supit{f},
D. Ward-Thompson\supit{o}
\skiplinehalf
\supit{a}Department of Physics \& Astronomy, University of Pennsylvania, 209 South 33rd Street, Philadelphia, PA, 19104, U.S.A.;\\
\supit{b}Cardiff University, School of Physics \& Astronomy, Queens Buildings, The Parade, Cardiff, CF24 3AA, U.K.;\\ 
\supit{c}Department of Astronomy \& Astrophysics, University of Toronto, 50 St. George St., Toronto, ON M5S 3H4, Canada;\\
\supit{d}Center for Interdisciplinary Exploration and Research in Astrophysics -- Northwestern University, 2145 Sheridan Road, Evanston, IL, 60208, U.S.A.;\\
\supit{e}Graduate School of Science, Nagoya University, Nagoya, Aichi 464-8602, Japan;\\
\supit{f}Department of Physics, Brown University, 182 Hope Street, Providence, RI, 02912, U.S.A.;\\
\supit{g}Department of Physics \& Astronomy, Northwestern University, 2145 Sheridan Road, Evanston, IL, 60208, U.S.A.;\\
\supit{h}California Institute of Technology, 1200 E. California Blvd., Pasadena, CA, 91125, U.S.A.;\\
\supit{i}Department of Physics, University of Toronto, 60 St. George St., Toronto, ON M5S 1A7, Canada;\\
\supit{j}Canadian Institute for Advanced Research, 180 Dundas St. W., Suite 1400, Toronto, ON M5G 1Z8;\\
\supit{k}Instituto de Astrof\'isica de Canarias, E-38200 La Laguna, Tenerife, Spain;\\
\supit{l}Universidad de La Laguna, Dept. Astrof\'isica, E-38206 La Laguna, Tenerife, Spain;\\
\supit{m}Department of Physics \& Astronomy, University College London, Gower Street, London, WC1E 6BT, U.K.;\\
\supit{n} Institute d'Astrophysique Spatiale, Bat 120-121, 91405 Orsay, France;\\
\supit{o}Jeremiah Horrocks Institute, University of Central Lancashire, PR1 2HE, U.K.;\\
}

\authorinfo{Nicholas Galitzki: E-mail: galitzki@sas.upenn.edu, Telephone: (215) 573-7558\\
Copyright 2014 Society of Photo-Optical Instrumentation Engineers. One
print or electronic copy may be made for personal use only. Systematic
reproduction and distribution, duplication of any material in this paper
for a fee or for commercial purposes, or modification of the content of
the paper are prohibited.}

\begin{document} 
  \maketitle 

\begin{abstract}
The Balloon-borne Large Aperture Submillimeter Telescope for Polarimetry (BLASTPol) is a suborbital mapping experiment, designed to study the role played by magnetic fields in the star formation process. BLASTPol observes polarized light using a total power instrument, photolithographic polarizing grids,  and an achromatic half-wave plate to modulate the polarization signal. During its second flight from Antarctica in December 2012, BLASTPol made degree scale maps of linearly polarized dust emission from molecular clouds in three wavebands, centered at 250, 350, and $500\, \mu$m. The instrumental performance was an improvement over the 2010 BLASTPol flight, with decreased systematics resulting in a higher number of confirmed polarization vectors. The resultant dataset allows BLASTPol to trace magnetic fields in star-forming regions at scales ranging from cores to entire molecular cloud complexes.
\end{abstract}


\section{INTRODUCTION}
\label{sec:intro}  

\input{gondola}

The Balloon-borne Large Aperture Submillimeter Telescope for Polarimetry (BLASTPol)\cite{Fissel2010} is a 1.8-meter Cassegrain telescope with three bolometric arrays operating over 30\% bandwidths centered on 250, 350, and 500 $\mu$m, which have 139, 88, and 43 bolometric detectors with diffraction limited resolution of 30\arcsec, 42\arcsec, and 60\arcsec, respectively. It was flown on a stratospheric balloon platform as part of NASA's Long Duration Balloon (LDB) program. BLASTPol detects linearly polarized emission from dust with a thermal emission peak in the submillimeter. The instrument flight configuration is show in in Figure \ref{fig:gondola}.

The detectors were originally developed as part of the SPIRE instrument program for {\it Herschel}\cite{Griffin2010} and were adapted for use on the first BLAST instrument, which was in turn modified to be sensitive to polarization. BLAST\cite{Pascale2008} has flown twice on LDB missions\cite{Marsden2008}\cite{Truch2009} which generated a legacy of important scientific results\cite{Devlin2009,Marsden2009,Pascale2009,Patanchon2009,Viero2009,Wiebe2009,Netterfield2009}. BLASTPol has also flown once before from Antarctica in 2010\cite{Pascale2012}, which proved the viability of mapping polarized dust in the submillimeter and provided the motivation for a second flight in 2012 to observe additional science targets. 

In order to make BLAST polarization sensitive, a polarizing grid was placed at the entrance to each feedhorn on all three arrays and a stepped Achromatic Half Wave Plate (AHWP)\cite{Moncelsi2014} was added to the optical configuration. The primary science goal of the new instrument, BLASTPol, was to probe polarized dust emission in Galactic star forming regions, in order to detect the local magnetic field orientation as projected on the sky. The polarization angular dispersion can also be used to estimate magnetic field strength when combined with velocity dispersion obtained from the width of molecular emission lines\cite{Chandrasekhar1953}.

During the second flight of BLASTPol from Antarctica in 2012 we made degree-scale maps of a number of nearby molecular clouds. We were able to obtain a vast improvement on the number of polarization pseudo-vectors over the 2010 flight. The 2010 instrumental performance was limited by a melted IR blocking filter that greatly complicated polarization analysis and reduced the resolution to $\sim$2.5\arcmin $\,$after smoothing with a Gaussian kernel\cite{Matthews2014}. A significant portion of the data from 2010 was also rejected due to contamination by intermittent systematic noise, which effectively reduced integration time on the targets. Our preliminary polarization analysis of the 2012 data indicates polarization levels of a few percent, with many of our pseudo-vectors noticeably correlated with molecular cloud structure. 

Data from the BLASTPol 2012 flight will allow us to fully explore the potential of this type of instrument to investigate the effects of magnetic field on cloud structure, from large filaments and extended regions into prestellar cores. The flight has delivered on the goals of improving the area, depth, and resolution covered, in comparison with the 2010 flight. The analysis will be able to show the effectiveness of this type of experiment in linking large-scale {\it Planck} polarimetry maps to the small scale, but high resolution, polarimetry maps observed with interferometers such as ALMA. 

\section{Magnetic Fields in Star-Forming Regions}
\label{sec:magfield}  

An important goal of modern astrophysics is to understand the star formation process and especially the factors that regulate the star formation rate (SFR) in molecular clouds and galaxies as a whole. Recent progress includes using observations of dust emission and extinction, which shows how core mass distribution correlates with observed stellar mass distribution\cite{Nutter2007}. There are also many results from the {\it Herschel} data. One example is evidence that filamentary structures are common in molecular clouds and that higher density filaments preferentially break into pre-stellar cores\cite{Andre2010,Hill2011}. However, there are many questions concerning the star-formation process and the evolution of cloud structure that remain to be addressed\cite{McKee2007}. Examples include whether the lifetimes of molecular clouds and their internal structures are equal\cite{Vazquez2006} to or larger\cite{Netterfield2009,Blitz2007,Goldsmith2008} than the turbulent crossing time. To have lifetimes longer than crossing times would require a supporting mechanism to counteract gravity. Magnetic fields could provide such support and numerical simulations have shown magnetic fields in clouds can drastically alter star formation efficiencies and the lifetimes of molecular clouds\cite{Li2010,Hennebelle2011}.  However, knowledge of magnetic fields and their interaction with molecular cloud structure is still fairly limited. Zeeman splitting observations have produced measurements of the field strength along the line of sight, but are limited to bright regions, and optical extinction polarization observations have produced a small number of magnetic field pseudo-vectors,  but only in areas of low extinction\cite{Crutcher2010,Falgarone2008}. The most promising method for detecting magnetic fields over large ranges of dust column density is with far-IR and submillimeter polarimetry\cite{Hildebrand2000,WardThompson2000,WardThompson2009}. Spinning dust grains preferentially anti-align with the local magnetic field and emit modified blackbody radiation with peaks in the 100s of $\mu$m range that is polarized orthogonally to the local magnetic field. BLASTPol is the first instrument capable of creating degree-scale polarization maps of molecular clouds with sub-arcminute resolution and a mapping speed that has allowed it to cover multiple targets during each flight. BLASTPol data enables direct comparison between polarization maps and numerical simulations\cite{WardThompson2000} and shows early agreement with previous observations of molecular cloud polarization\cite{WardThompson2009,Li2006}. 

BLASTPol observations target the following three key questions in star formation (further discussed in Ref. \citenum{Fissel2010}):\emph{
i) Is core morphology and evolution determined by large-scale magnetic fields? ii) Does filamentary structure have
a magnetic origin? iii) What is the field strength, and how does it vary from cloud to cloud?}

\section{Instrument} 
\label{sec:instrument} 

BLASTPol was created by modifying the BLAST instrument, which is described in detail in Refs. \citen{Fissel2010,Pascale2008,Marsden2008}. The optical layout is described in Figure \ref{fig:optics} with the optical component parameters listed in Table \ref{tab:optica}. The telescope uses a Ritchey-Chr\'etien configuration with an aluminum 1.8-m diameter primary mirror attached to a 40-cm aluminum secondary mirror. The secondary mirror is actuated to allow for active focusing during the flight to adjust to the differential thermal contraction of the support structure. The beam is then reimaged by the cold optics ($\sim$1.5 K), which are situated in an Offner relay configuration. The light is then split by two dichroic filters \cite{Ade2006} into the science bands at 250, 350, and 500 $\mu$m. The focal planes are held within a cryostat cooled by liquid helium and liquid nitrogen, which has a hold time of approximately 13 days. The total field of view (FOV) of each array is 14\arcmin x 7\arcmin. The arrays are kept at 290 mK by a closed-cycle \textsuperscript{3}He fridge. Each pixel is made of a resistance thermal device (RTD) bolometer glued to a silicon-nitride ``spider-web''\cite{Bock1998} absorptive element, which is coupled to a smooth-walled conical feedhorn\cite{Chattopadhyay2003} spaced at $2f\lambda$ .

The primary scanning strategy during flight is a slow raster scan, which works well for extended sources. A typical BLASTPol raster scans across targets in azimuth at a speed of $\sim 0.1^\circ{\rm s}^{-1}$ with an elevation scan speed calculated to change the elevation by 1/3 the array FOV in one crossing.
\input{optics}

\input{tab_optic_v2}

Part of the characterization of instrumental performance is carried out using near-field mapping of the beam profile with the fully assembled optics system. These tests are performed by placing a chopped liquid nitrogen cooled source 100 m away from the telescope on a stage that can be moved in a plane parallel to the surface of the mirror, in order to sample over the whole array. To enable the telescope to view a near-field source the secondary mirror is offset away from the primary mirror by $\sim$3 cm using aluminum blocks that are inserted into the mounting point of the secondary mirror to its support struts. Fine tuning of the focus is accomplished with the active focus actuators on the secondary mirror. Due to atmospheric absorption we are only able to effectively determine the beam shape at 500 $\mu$m. During the course of the 2010 campaign a noticeable lobe, offset from the primary beam by $\sim$1\arcmin, with $\sim$30\% of the total power was observed during mapping of the primary beam. A detailed analysis prior to the 2012 flight revealed that the weight of the secondary mirror was distorting the primary mirror at the strut mounting points on the rim of the mirror. This problem was addressed by shifting the mounting points of the secondary mirror to the telescope support frame structure. A detailed description of the configuration and testing of the instrument for the 2010 and 2012 flights will be given in Angil\'e et al. (2014, in preparation).

\input{polgrids}

\section{Polarimetry} 
\label{sec:polarimetry} 

Each detector array has a photolithographed linear polarizer (Figure \ref{fig:polgrids}) mounted to the front of the feedhorn block. The polarizing grid orientation rotates by $90^\circ$ from one pixel to the next along each row. The rows are parallel to the nominal scan direction. This alignment allows for sampling of either a {\it Q} or {\it U} Stokes parameter on a timescale that is much shorter than the array's common mode $1/f$ noise, which has a knee at 0.035 mHz\cite{Pascale2008}. The sampling timescale of the Stokes parameter is $\sim$0.125 s, which is determined by the detector separation, 45\arcsec  at 250 $\mu$m, and typical scan speed, $\sim 0.1^\circ{\rm s}^{-1}$. The field at the end of each feedhorn is approximately Gaussian, which results in very small leakage in polarization between adjacent pixels, estimated to be less than 0.07\% \cite{Moncelsi2011b}.

The AHWP is used to modulate the polarization signal, so that each pixel samples {\it I}, {\it Q}, and {\it U} multiple times during the course of a scan. The AHWP used in BLASTPol\cite{Moncelsi2014} has a 10-cm diameter aperture and is made from five layers of 500 $\mu$m thick sapphire, which are glued together with a 6 $\mu$m layer of polyethylene. The outer faces of the AHWP have a metal-mesh anti-reflective (AR) coating \cite{Zhang2009}. 

The AHWP is driven by a gear-train connected by a G-10 shaft to a stepper motor mounted on the exterior of the cryostat. The shaft is sealed via a vacuum ferromagnetic feedthrough\footnote{Ferrotech Corporation: 526 S. Jefferson St., New Castle, PA 16101}. The AHWP is stepped between four set angles (0$^\circ$, 22.5$^\circ$, 45$^\circ$, and 67.5$^\circ$) after each completed scan of a source in elevation. The position is determined by an absolute reading from a 4-K potentiometer on the rim of the AHWP mount, combined with the stepper-motor encoder. Due to some parts of the potentiometer reading poorly in pre-flight tests, the positions used in the 2012 flight were offset from the positions in the 2010 flight by 15.12$^\circ$. A detailed description of the AHWP design and performance is included in Ref. \citen{Moncelsi2014}. 

We conducted pre-flight ground tests to characterize the polarization properties of the instrument, using a chopped heated source placed directly in front of the receiver window, in order to fill all pixels. In one test the source was unpolarized and the instrumental polarization (IP) signal was observed. The IP was determined to be less than 1\% for all three arrays. In a second test we polarized the source with a polarizing grid tilted at 45$^\circ$ with respect to the incident beam and measured the polarization efficiency (PE) of the system. This was observed to be 80\%, 77\%, and 85\% for the 250, 350, and 500 micron arrays, respectively. Lower PE at shorter wavelengths is expected, due to the especially large spectral range observed\cite{Moncelsi2014} and the fact that the instrument was optimized for performance at 500 $\mu$m, as motivated by scientific considerations.

\section{Flights} 
\label{sec:flights} 

\input{tab_obs2012_v2}

The BLAST program has flown LDB missions four times. In 2005 BLAST flew from Kiruna, Sweden and in 2006 from McMurdo, Antarctica. These flights resulted in a number of ground breaking results, notably: measurements of the FIR background\cite{Devlin2009,Marsden2009,Pascale2009,Patanchon2009,Viero2009,Wiebe2009,Netterfield2009} and studies of the galaxies that comprise it; and a high resolution map of the Vela C molecular cloud\cite{Netterfield2009}, which reinforced the case for support mechanisms that reduce star formation rates. A proposed mechanism is trapped magnetic field lines that can be observed through polarized dust emission in the submillimeter. BLASTPol was created primarily to investigate this phenomenon and subsequently flew from Antarctica in 2010 and 2012. The first flight produced novel results\cite{Matthews2014,Poidevin2014}, despite a large IP caused by a melted IR filter. 

The 2012 flight was considerably more successful, producing thousands of pseudo-vectors on a number of important targets, despite hard-drive failures in our star cameras. One camera failed about 6 hours after launch and the second failed after 6 days, which has made the pointing reconstruction for the latter half of the flight challenging. The data from the last 6 days will consequently take longer to process. Our results indicate polarization fractions of a few percent in many regions, significant correlation with cloud structure, and polarization spectra that will serve to constrain dust grain models. A summary of the observed targets is given in Table \ref{tab:obs2012}.

Our observations of magnetic field direction indicate strong agreement with near-IR polarization measurements in the Vela C molecular cloud. 
We also see alignment between pseudo-vectors in our three observational bands. We expect to see similarly exciting results in the analysis of the other fields we have observed. The success of the 2010 and 2012 missions has led to the development of BLAST--The Next Generation (BLAST-TNG)\cite{Dober2014}, an entirely new instrument with a larger mirror and kinetic inductance detector (KID) arrays with a planned launch from Antarctica in December 2016. This new instrument will be able to observe many more sources and produce $\sim$16 times more polarization vectors than BLASTPol.

\section{Data Reduction and Preliminary Results} 
\label{sec:data} 

The data reduction pipeline is based on the original BLAST analysis methods, along with additional methods developed to analyze polarization in the 2010 BLASTPol data. The pipeline is described in detail in Refs. \citen{Pascale2008,Patanchon2009}. The first steps in the analysis include masking contaminated data from cosmic rays, glitches, and noise from the Tracking and Data Relay Satellite System (TDRSS) antenna. The TDRSS contamination was recognized early in the flight and only affected a small percentage of the data. After pre-processing, detectors are then corrected for drifts in responsivity, using the signal from a calibrator lamp that is flashed every $\sim$15 minutes. Common-mode detector response from changes in elevation pointing are also removed, along with other artifacts inherent to the data, at which point the arrays are flat-fielded relative to a chosen pixel, using aperture photometry from one of our calibration targets.

IP is determined by examining maps made of the same polarized source at different sky rotations. The IP is then subtracted out and the data are processed by an ideal map maker, which is under development. The map maker produces Stokes {\it I}, {\it Q}, and {\it U} maps from which a polarization pseudo-vector and polarization fraction can be extracted. The direction of the vector is then rotated by $90^\circ$ to get the correct orientation of the magnetic field. These data are then analyzed using a variety of simulations and jackknife consistency checks to determine uncertainty levels of the polarization strength and orientation. Pseudo-vectors that do not pass these tests are rejected. This process is described in detail in Ref. \citen{Matthews2014}. We plan to publish results from the analysis within the next six months.

\section{Conclusion} 
\label{sec:conclusion} 

The viability of a balloon-borne telescope to explore the polarization of the submillimeter sky has been shown with the success of the BLASTPol 2012 Antarctic flight in achieving its primary science goals. We have mapped a number of important Galactic targets with an unprecedented combination of resolution and sky coverage. For the first time we have magnetic field maps that span entire molecular cloud structures that will link the full sky polarimetry maps of {\it Planck} with the high resolution, small area polarimetry maps, of telescopes such as the SMA and ALMA. This will strongly constrain models of magnetic field and turbulent interaction within magnetic fields as well as constraining dust grain models. 

The BLAST and BLASTPol experiments have completed a $\sim$8 year legacy of important scientific results and advancement of balloon borne telescopes. This has provided a solid foundation for the next generation of instruments. The readout\cite{Benton2014}, pointing\cite{Gandilo2014}, and thermal systems\cite{Soler2014} that have been advanced during the course of this experiment will continue to see use on a number of other instruments and BLASTPol's success has led directly to the development of the BLAST-TNG project.

\section{Acknowledgments} 
The BLAST collaboration acknowledges the support of NASA through grant numbers NNX13AE50G S03 and NNX09AB98G, the Canadian Space Agency (CSA), the Leverhulme Trust through the Research Project Grant F/00 407/BN, Canada's Natural Sciences and Engineering Research Council (NSERC), the Canada Foundation for Innovation, the Ontario Innovation Trust, the Rhode Island Space Grant Consortium, and the National Science Foundation Office of Polar Programs. F. Poidevin thanks the Spanish Ministry of Economy and Competitiveness (MINECO) under the Consolider-Ingenio project CSD2010-00064 (EPI: Exploring the Physics of Inflation) for its support. J. D. Soler acknowledges the support the European Research Council under the European Union's Seventh Framework Programme FP7/2007-2013/ERC grant agreement number 267934. We would also like to thank the Columbia Scientific Balloon Facility (CSBF) staff for their continued outstanding work. 


\bibliography{blast}   
\bibliographystyle{spiebib}   

\end{document}

%% file: gondola.tex
\begin{figure*}[t]
  \centerline{
    \includegraphics[height=3.5in,natwidth=1681,natheight=897]{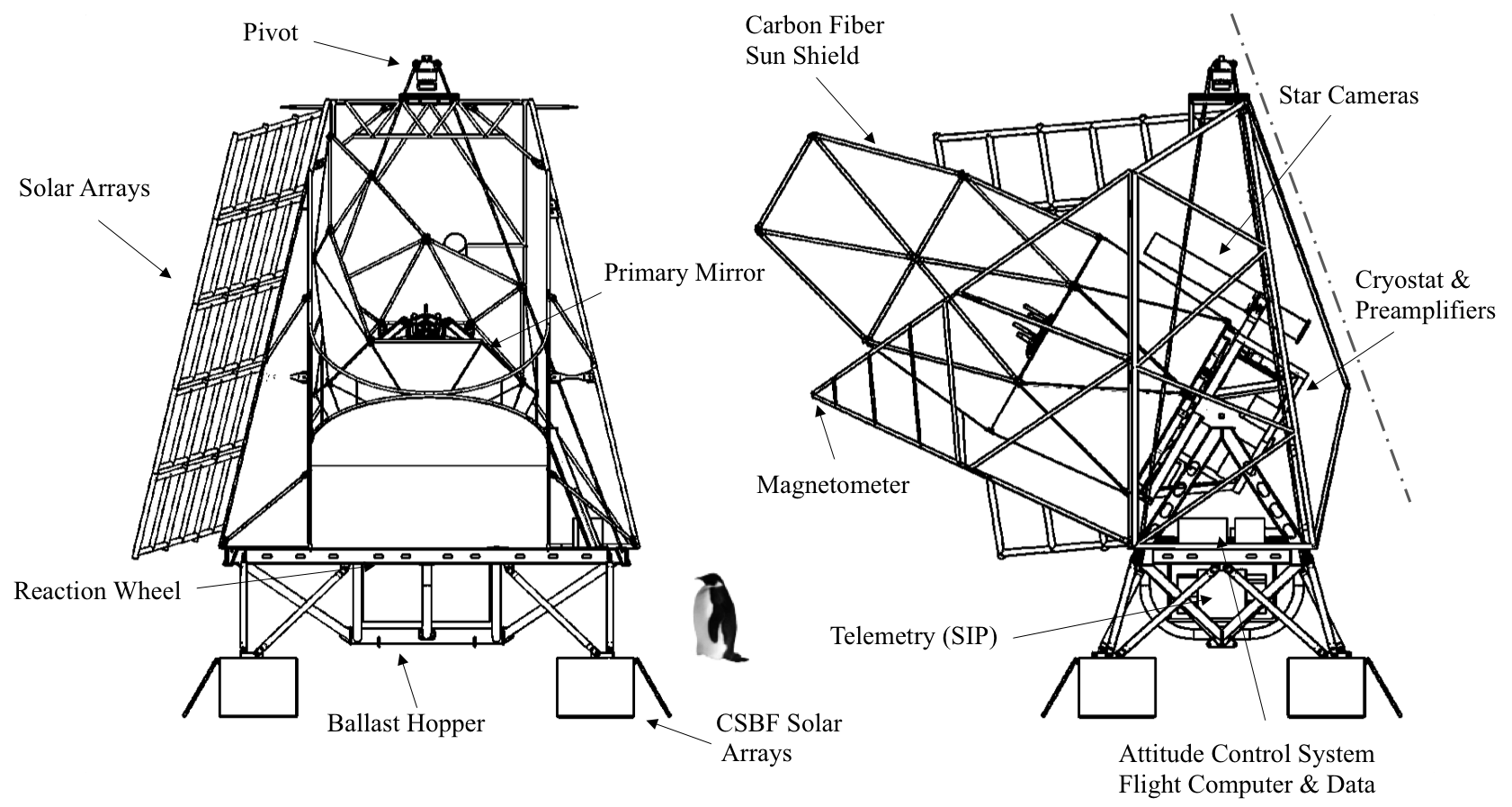}
  }
  \caption{The front and side profile views of the BLASTPol gondola. The cryostat, primary mirror, secondary mirror struts, star cameras, and detector readout electronics are mounted to an inner frame that points in elevation. An asymetric carbon fiber sunshield is also attached to the inner frame and allows the telescope to point within 45$^\circ$ of the sun in azumith. The 1-m tall Emperor penguin is shown for scale. 
    \label{fig:gondola}}
\end{figure*}

%% file: optics.tex
\begin{figure}[t]
  \centerline{
    \includegraphics[height=2.5in]{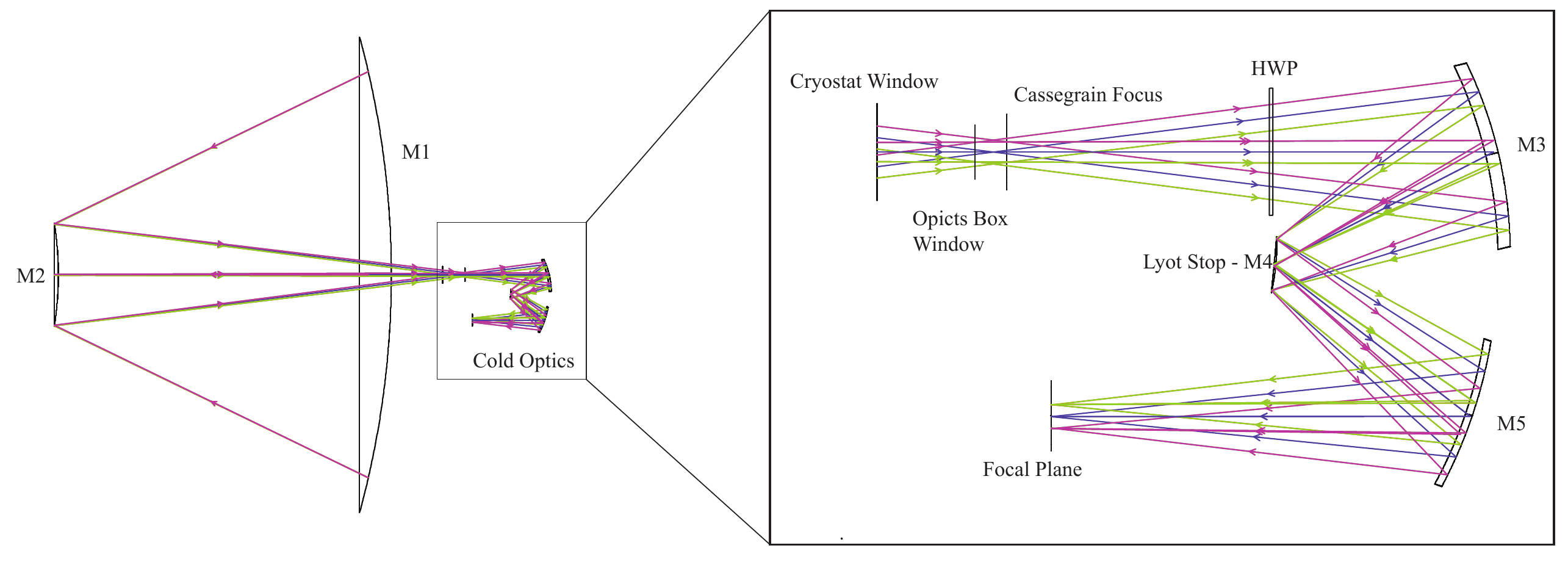}
  }
  \caption[optics]
	{\label{fig:optics} Side view of the optical layout of the BLASTPol telescope and receiver with a detailed view of the cold, 1.5-K, optics that are internal to the cryostat. The cold optics mirrors are in an Offner configuration with spherical mirrors M3, M4, and M5. M4 acts as a Lyot stop, which defines the illumination of the primary mirror for the arrays. Two dichroic filters(not shown) separate the beam into the three science bands and are placed between M5 and the $500\,\mu$m focal plane. The AHWP is mounted between the Cassegrain focus and M3.}
\end{figure}

%% file: tab_optic_v2.tex
\begin{table}[t]

\newcommand\T{\rule{0pt}{3ex}}       
\newcommand\B{\rule[-2ex]{0pt}{0pt}} 
\vspace{3 mm}
\begin{center}
\caption{Summary of BLASTPol Optics Characteristics\label{tab:optica}}
\vspace{3 mm}
\renewcommand{\arraystretch}{1.2}%
\begin{tabular}{lccccc}
\hline
\hline
\T Geometrical Charac. & M1 & M2 & M3 & M4 & M5 \B \\
\hline
\T Nominal Shape & Paraboloid & Hyperboloid & Sphere & Sphere & Sphere\\                                                       
Conic Constant & $-$1.029 & $-$2.853 & 0.000 & 0.000 & 0.000\\
Radius of Curvature & 4.186$\,$m & 1.154$\,$m & 348.6$\,$mm & 174.3$\,$mm & 348.8$\,$mm\\
Aperture & \diameter1.816$\,$m & \diameter0.399$\,$m & 95$\times$ 75$\,$mm & \diameter36.8$\,$mm & 95$\times$ 75$\,$mm \B \\
\hline
\end{tabular}
\end{center}
\end{table}

%% file: polgrids.tex
\begin{figure}[t]
\begin{center}
\includegraphics[height=3in]{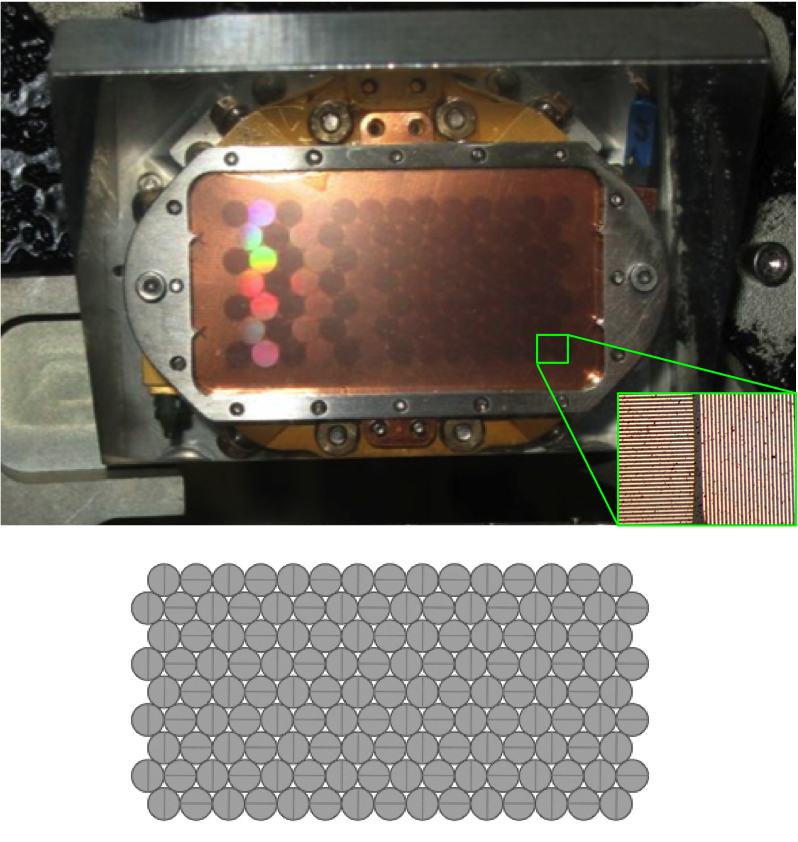}
\end{center}
\caption[polgrids]
{\label{fig:polgrids}
Top: An image of the $350\,\mu m$ array's photolithographed polarizing grid with an inset showing the grid alternating by 90$^\circ$ between pixels. Bottom: A figure that depicts the directions of the grid at each pixel. The layout allows for measurement of a {\it Q} or {\it U} Stokes parameter on timescales $<$ 1 second.}
\end{figure}

%% file: tab_obs2012_v2.tex
\begin{table}[t]

\newcommand\T{\rule{0pt}{3ex}}       
\newcommand\B{\rule[-2ex]{0pt}{0pt}} 
\vspace{3 mm}
\begin{center}
\caption{BLAST-Pol 2012-2013 Observed Primary Targets\label{tab:obs2012}}
\vspace{3 mm}
\begin{tabular}{llccccccc}
\hline
\hline
\multicolumn{1}{c}{Target} \T & \multicolumn{1}{c}{Type} & Distance & Size & Expected No. of B-vectors \\
 &  & (pc) & (deg$^{2}$) & \B\\ 
\hline
\T Vela C & Nearby GMC & $\sim$700 & 12 & $\sim$8000 \\
Carina Nebula & Calibrator (GMC) & $\sim$2300 & 2 & $\sim$3000 \\
CG12 & Low Mass Cloud & $\sim$550  & 0.1 & TBD \\
G331 & Calibrator (GMC) & $\sim$7000 & 2 & $\sim$4000 \\
IRAS 08470-4243 & Calibrator & $\sim$700 & 0.1 & NA \\
Lupus I & Dark Cloud & $\sim$155 & 1 & $\sim$200 \\
Puppis Cloud Complex & Nearby Cloud & $\sim$1900 & 0.4 & TBD \B \\
\hline
\end{tabular}
\end{center}
\end{table}